# The chemical composition of CO-rich comet C/2009 P1 (Garradd) at $R_h$ = 2.4 and 2.0 AU before perihelion[1]

L. Paganini[1,2], M.J. Mumma[1], G.L. Villanueva[1,3], M.A. DiSanti[1],

B.P. Bonev[1,3], M. Lippi[4], and H. Boehnhardt[4]

[1] Goddard Center for Astrobiology, NASA GSFC, MS 690, Greenbelt, MD 20771, USA

[2] NASA Postdoctoral Fellow, lucas.paganini@nasa.gov

[3] Department of Physics, Catholic University of America, Washington, DC 20064, USA

[4] Max-Planck-Institut für Sonnensystemforschung, Katlenburg-Lindau, Germany

[1] Based on observations obtained at the European Southern Observatory at Cerro Paranal, Chile, under program 087.C-0710.

## Abstract

We quantified ten parent volatiles in comet C/2009 P1 (Garradd) before perihelion, through high-dispersion infrared spectra acquired with CRIRES at ESO's VLT on UT 2011 August 07 ($R_h$ = 2.4 AU) and September 17–21 ($R_h$ = 2.0 AU). On August 07, water was searched but not detected at an upper limit (3σ) of $2.1 \times 10^{28}$ $s^{-1}$, while ethane was detected with a production rate of $6.1 \times 10^{26}$ $s^{-1}$ (apparent mixing ratio > 2.90%). On September 17-21, the mean production rate for water was $8.4 \times 10^{28}$ $s^{-1}$, and abundance ratios (relative to water) of detected trace species were: CO (12.51%), $CH_3OH$ (3.90%), $CH_4$ (1.24%), $C_2H_6$ (1.01%) and HCN (0.36%). Upper limits (3$\sigma$) to abundances for four minor species were: $NH_3$ (1.55%), $C_2H_2$ (0.13%), HDO (0.89%) and OCS (0.20%). Given the relatively large heliocentric distance, we explored the effect of water not being fully sublimated within our FOV and identified the 'missing' water fraction needed to reconcile the retrieved abundance ratios with the mean values found for "organics-normal". The individual spatial profiles of parent volatiles and the continuum displayed rather asymmetric outgassing. Indications of $H_2O$ and CO gas being released in different directions suggest different active vents and/or the possible existence of polar and apolar ice aggregates in the nucleus. The high fractional abundance of CO identifies comet C/2009 P1 as a CO-rich comet.

## 1. Introduction

Until 2005, cometary nuclei were thought to be primordial remnants from the giant planets' accretion zone (then viewed as 5–30 AU from the proto-Sun), but subsequent dynamical studies challenge this view and demonstrate the need for an alternative metric of cometary origins (Morbidelli et al. 2008; Levison et al. 2010; Walsh et al. 2011). Those origins are most directly inferred from the properties of native materials (dust, ice) in the nucleus.

Since 1985, measurements of nucleus composition have revealed major diversity amongst comets, based on crystallinity of silicates and/or the chemistry of native ices. The interpretation, however, is partly linked to questions of orbital evolution. A comet making its first apparition after ejection from the long-term storage reservoir (Kuiper Belt, Oort Cloud) may reveal primordial composition (although cosmic ray processing might affect the properties of the outer layers of cometary nuclei during dynamical storage). Comets in short period orbits, on the other hand, may have experienced thermo-chemical evolution over successive apparitions that may induce changes from primordial composition. In addition, the number of comets quantified to date in terms of primary (parent) volatiles is still relatively small, restricting full taxonomic classification of origins (for a recent review of cometary taxonomies and natal heritage, see Mumma & Charnley 2011). Here, we present results for the volatile fraction of Oort Cloud (OC) comet C/2009 P1 (Garradd), hereafter C/2009 P1, and discuss them in the context of an emerging cometary taxonomy based on primary volatiles.

C/2009 P1 was discovered at $R_h$ = 8.7 AU by G. J. Garradd on UT 2009 August 13, when it displayed a circular (15″ diameter) coma with a visual magnitude ~17. Its orbit is inclined to the ecliptic by 106.2°, and the Tisserand parameter ($T_j$) and original semi-major axis[2] are -0.432 and 2564.1 AU, respectively. Together, these classify C/2009 P1 as a nearly isotropic (long period) comet from the OC reservoir.

In this Letter, we present results obtained from near IR observations of comet C/2009 P1 at the Very Large Telescope. These results, along with our current characterization of its organic-rich gas production and favorable astrometric positioning, establish C/2009 P1 as a prime target for astronomical observations through early 2012.

## 2. Observations

We observed comet C/2009 P1 with CRIRES at the Very Large Telescope (VLT) located in the Atacama desert (Chile), on UT 2011 August 07 and on five consecutive nights spanning September 17–21[3]. The Cryogenic high-Resolution InfraRed Echelle Spectrograph (CRIRES) provided a spectral resolution ($\lambda/\Delta\lambda$) of about 50,000 using an entrance slit of 0.4″ (in width) and a spatial coverage of 40″ (in length) (Käufl et al. 2004). Weather conditions were optimal during our September observations, with low wind speeds of 3 m $s^{-1}$ (5 m $s^{-1}$), relative humidity of 5% (10%) and water vapor of 1.2 mm (2.1 mm) on the first two nights (remaining nights). Seeing was in the range 0.5″ – 0.9″. Standard stars (BS 7235 and BS 7906) located near the comet's trajectory allowed flux calibration and measures of column burdens for absorbing species in the terrestrial atmosphere. The CRIRES entrance slit (0.4″ × 40″) was positioned along the projected Sun-comet direction on all five nights. The observing log is given in Table 1.

---

[2] S. Nakano Note, NK 2109, http://www.oaa.gr.jp/~oaacs/nk/nk2109.htm

[3] The core of this paper is based on the September dates. During the August run, there were difficulties to close AO loop (limiting the sensitivity of our observations). Results from UT 2011 August 07 are presented briefly in Section 3.2, Tables 1 and Fig. 3.

Cometary spectra were acquired in a four-step sequence (ABBA) with an integration time of 60 (or 120) seconds per step, nodding the telescope along the slit by 11″ between the A and B position. We followed our standard procedures for initial data reduction and analysis of the individual echelle orders (DiSanti et al. 2001; Villanueva et al. 2011a), which included flat fielding, removal of high dark current pixels and cosmic-ray hits, spatial and spectral rectification, and spatial registration of individual A and B beams. After combining the A and B beams from the difference frames, we extracted a spectrum from the sum of 15 spatial rows (1.29″) centered on the nucleus (this also assists in removing the residual background).

The simultaneous measurement of $H_2O$ or its direct proxy (OH*) in each instrument setting permitted accurate characterization of the abundance ratio for each trace species. Six primary volatiles ($H_2O$, CO, $C_2H_6$, $CH_4$, HCN, $CH_3OH$) were securely detected, and upper limits were retrieved for $NH_3$, $C_2H_2$, OCS, and HDO. As mentioned in our IAU Circular (Paganini et al. 2011) and shown in Figure 1, detections of eight prominent CO emission lines enabled robust retrieval of a (high) $CO/H_2O$ ratio. Detections of other primary volatiles are shown in Figure 2. Several unidentified lines are seen in these spectra, however we defer a complete analysis to a future publication. Figure 3 shows the temporal development of specific production rates. Figure 4 displays spatial profiles for six volatiles, the cometary continuum and the flux standard star for all days. On the third night (Sept 19), we emphasized OCS and HDO but obtained only upper limits, and a retrieval of spatial profiles was not possible.

## 3. Results and discussion

Comet C/2009 P1 displayed significant outgassing at these pre-perihelion positions ($R_h$ = 2.4 and 2.0 AU). Specific results are shown in Table 1 for each setting, date, and volatile species. In the following subsections we highlight our major findings.

### 3.1. Rotational temperatures

We retrieved accurate rotational temperatures for $H_2O$, CO, $C_2H_6$, HCN, and $CH_3OH$ by performing excitation analyses of extracted and predicted intensities for spectral lines of a given molecule. Predicted intensities were obtained from our custom fluorescence models (Villanueva 2011b; Villanueva 2011a, and references therein). A rotational temperature of 50 K is consistent with the average temperature from these five volatiles, although some minor differences are observed (Table 1).

### 3.2. Production rates and abundance ratios

Production rates of $H_2O$ and HCN displayed a rather stable outgassing during the five nights in September, as opposed to $C_2H_6$, CO, and $CH_4$ whose relative production decreased by ~15% from Sept 18–19 to 21. During this interval, $H_2O$ and HCN changed by less than 5% (see Fig. 3). The mean production rates (over the entire observing interval) for each volatile species are: $H_2O$ ($8.4 \times 10^{28}$ $s^{-1}$), CO ($1.1 \times 10^{28}$ $s^{-1}$), $CH_3OH$ ($3.3 \times 10^{27}$ $s^{-1}$), $CH_4$ ($1.0 \times 10^{27}$ $s^{-1}$), $C_2H_6$ ($8.5 \times 10^{26}$ $s^{-1}$), and HCN ($3.0 \times 10^{26}$ $s^{-1}$). Abundances relative to water are: CO (12.51%), $CH_3OH$ (3.90%), $CH_4$ (1.24%), $C_2H_6$ (1.01%), and HCN (0.36%). Upper limits ($3\sigma$) were obtained for the abundance ratio for four minor species: $NH_3$ (1.55%), $C_2H_2$ (0.13%), HDO (0.89%), and OCS (0.20%).

On 2011 August 07 ($R_h$ = 2.4 AU), water was searched but not detected at an upper limit ($3\sigma$) of $2.1 \times 10^{28}$ $s^{-1}$, while ethane was detected with a production rate of $6.1 \times 10^{26}$ $s^{-1}$. When compared with measurements of $C_2H_6$ in mid-September, the production of ethane in August is consistent with insolation-limited outgassing (varying as $R_h^{-2.2}$). The upper limit for water suggests a mixing ratio for ethane > 2.9% (3σ), which far exceeds the abundance ratio measured in late September. The likely explanation is that water sublimation was less fully activated at 2.4 AU, compared with 2.0 AU.

D. Schleicher 2011 (private communication) reports water production rates at ($R_h$) 2.49 AU (July 28) and 1.97 AU (Sept. 24) that are larger than our values on nearby dates. If we adjust the sublimated water fraction in mid-September to achieve "normal" abundances of $C_2H_6$ (~ 0.52%) and/or HCN (~ 0.23%), the water production rate would increase by about 60–90%, resulting in abundance ratios of CO ~ 7.26%, $C_2H_6$ ~ 0.59%, $CH_4$ ~ 0.72%, HCN ~ 0.21%, and $CH_3OH$ ~ 2.26%. If the 'missing' fraction were pure water ice, this increase would reconcile water production rates estimated in the infrared to those obtained in the optical. An even larger 'missing' fraction is required at 2.4 AU, but *AKARI* observed an increase in mixing ratios for $CO_2$ beyond 2.5 AU (Ootsubo et al. 2011), demonstrating the decreasing efficiency of water ice sublimation at heliocentric distances beyond 2.5 AU.

### 3.3. Spatial profiles

During our September observations we oriented the slit along the extended comet-Sun projected radius vector (Position Angle, P.A., ~ 96º). The Solar Phase angle of 29.5º (Table 1) indicates that the sub-solar point was well placed within the nucleus hemisphere seen from Earth.

During the first night (UT 2011 Sept 17, see Fig. 4a), the (asymmetric) water profile displayed strong enhancement in the sunward direction (here, we use "sunward" to mean the directed projection on the sky plane of the comet-sun vector). The profile of ethane was also enhanced in the sunward direction, but less so than that of water. The methane profile was symmetric about the nucleus, and displayed a steep slope in both directions with a secondary minimum at ~ 2000 km sunward.

On the second night (UT 2011 Sept 18, see Fig. 4b), CO and HCN showed profiles that were somewhat enhanced in the anti-sunward direction. HCN displayed a secondary minimum at 2000 km sunward, similar to $CH_4$ on the previous night. Likewise, the profile of OH* (a direct tracer of $H_2O$) showed a behavior similar to that of water on the previous night.

Compared to the continuum, $CH_3OH$ showed some enhancement on both sides during the fourth night (UT 2011 Sept 20, see Fig. 4c), meanwhile HCN displayed flux excess toward negative pixels only (to the left in Fig. 4c). HCN displayed a broader profile than that seen on Sept 18, perhaps suggesting enhanced gas release on Sept 20. (A profile for $H_2O$ or OH* was not possible during this night due to the low signal-to-noise ratio of these lines.)

On the fifth night (UT 2011 Sept 21, see Fig. 4d), water showed a behavior similar to the first and second nights (i.e. some flux increase toward positive pixels). Ethane displayed a less asymmetric profile, compared with the first night, with minor flux excess in the anti-sunward direction. CO followed the continuum profile in the sunward direction, as opposed to the anti-sunward side, where its flux was clearly enhanced compared to the continuum (somewhat similar to HCN and $CH_3OH$ on the previous night).

The continuum profile displayed higher intensity toward the sunward direction on all nights (slope differences are confirmed between the anti- and sunward directions through log-log representations of the flux profile along the slit). These molecules displayed distinctive schemes of gas release, although some particular structures are found. Compared to the continuum profile, we notice that water (and OH*) is clearly enhanced in the sunward side on three nights (1, 2 and 5), similar to methanol on the fourth night. Conversely, spatial profiles of CO, HCN, and (also) $CH_3OH$ displayed flux excess in the anti-sunward direction; meanwhile $C_2H_6$ and $CH_4$ showed similar (rather symmetric) profiles. These configurations strengthen the idea of separate polar ($H_2O$-rich) and apolar (CO-rich) ice aggregates whose accreting gas composition underwent different chemical processes before incorporation in the nucleus, or re-distribution after incorporation.

The flux excess seen relative to the continuum for some volatiles (especially $H_2O$, CO and methanol) suggests the presence of a delayed sublimation mechanism, such as icy grains subliming in the coma. The spatial profiles of some volatiles display regular patterns, with a separation of ~1000–1500 km between peaks (see Fig. 4). These patterns are also present in observations of water emission lines with NIRSPEC in October 2011 (DiSanti et al. 2012, in preparation). On the other hand, active regions (jets) with heterogeneous composition and different rotational phases could also produce such behavior. This combination argues against icy grains being the only source of distributed release.

### 3.4. The (relatively) high abundance of CO

Measurements of cometary composition suggest rather low (or depleted) carbon monoxide relative to $H_2O$ in most comets (Bockelée-Morvan et al. 2004; Mumma and Charnley 2011). The measured abundance ratios (from ground-based studies) range from 0.2% to ~24%, but only four comets displayed CO > 10%, relative to water. A recent survey of 18 comets (10 JFCs and 8 OC comets) at heliocentric distances between 1 and 4 AU by the *AKARI* mission (Ootsubo et al. 2011), found a similar paucity of comets enriched in CO. Comets within 2.5 AU of the Sun revealed little (or no) CO content (mostly upper limits were retrieved), though their $CO_2$ abundance varied from a few to ~30% (relative to water). An exception was comet C/2008 Q3 (Garradd) that displayed $CO/H_2O$ ~ 30% (at $R_h$ = 1.7 AU). Abundance ratios of CO and $CO_2$ in comets surveyed beyond ~2.5 AU were likely "artificially" enhanced by the increasing stability (against sublimation) of water ice.

Valuable lessons are also taken from observations towards young stellar objects (YSOs). Ices control much of the star formation process and account for most oxygen and carbon in a protostellar environment (Öberg et al. 2011), so comparing these abundances to those found in comets provides useful tests of the possible influence of thermal (and chemical) processing of cometary material before accretion. Measurements in Galactic YSOs have resulted in $CO/H_2O$ in the range 2–20% (Chiar et al. 1998; Gibb et al. 2004) and $CO_2/H_2O$ 10–23% (Gerakines et al. 1999), which may indicate high ratios $CO_2/H_2O$ rather than depleted carbon monoxide in comets.

On the other hand, comets such as C/1995 O1 (DiSanti et al. 1999, 2001), C/1996 B2 (Mumma et al. 1996; Biver et al. 1999; DiSanti et al. 2003), C/1999 T1 (Mumma et al. 2003), C/2008 Q3 (Ootsubo et al. 2011), and C/2009 P1 (this work), revealed high abundances of CO relative to $H_2O$, and thus these "peculiar" exceptions (compared to the normal trend) are not trivial. Owing to its desorption temperature (15–30 K, depending on the ice mixture), CO ice is strongly susceptible to the effect of thermal processing by stellar radiation. A study by Shimonishi et al. (2010) confirmed the systematic difference in the CO ice abundance of luminous YSOs in the Large Magellanic Cloud (LMC), demonstrating the key influence of stellar radiation on CO ice abundance (unlike $CO_2$ ice).

Even though $H_2O$, CO and $CO_2$ are the most abundant ices in molecular clouds before the onset of collapse, their formation chemistry and content in comets are still poorly understood. Simulations of the chemical evolution considering the effect of turbulent transport of gases and ices in planetary disks have proven to be useful for understanding the measured abundances of these ices in comets (e.g. Semenov and Wiebe 2011). These results, along with observational evidence of clear diversity in the chemical composition of comets, demonstrate the importance of radial mixing and transport from different regions into the comet's formative zone (putatively, $R_h$ = 5–30 AU), probably at different evolutionary stages of the young protosun. Indeed, the long-held hypothesis relating crystallinity in cometary silicates to high-temperature processing of pre-cometary grains was recently confirmed by detailed laboratory investigations of rocky samples returned from 81P/Wild 2 by *Stardust* (Brownlee et al. 2006). Formed within 10 solar radii, these refractory silicates were carried outward to the region where even hypervolatile gases could condense, sheathing them in ices of water and organic volatiles. Successive aggregation later formed ever-larger bodies, culminating in the final icy planetesimal now known as the nucleus of comet Wild 2.

The possible connection between interstellar chemistry and cometary ices has been greatly strengthened by observations of several bright comets, especially C/1995 O1 (Hale-Bopp) (e.g., Ehrenfreund et al. 1997; Bockelée-Morvan et al. 2000). As Mumma and Charnley (2011) discussed in their review, planetesimals in cometary nuclei stem from interstellar matter that underwent different degrees of chemical and dynamical processing before aggregation. Thus, considering the formation of cometary nuclei from material with distinct chemical histories, i.e., various degrees of processing (including pristine material, cf. Visser et al. 2011), the high CO composition found in comet C/2009 P1 (and in other similar CO-rich comets) strengthens the idea that some material formed in outer regions in the disk, where stellar radiation was less intense and so CO was better shielded from external catalysts. Other possible scenarios include the possible capture of CO within the water ice mantles, which would produce desorption only at higher temperature. Perhaps the most extreme issue is the possible capture of these comets from stars in the Sun's birth cluster (Levison et al. 2010).

## 4. Conclusions

Among the organic compounds measured, we find a relative enhancement of organic volatiles in C/2009 P1 (Garradd), compared to other OC comets in our IR survey. Among the as-measured abundance ratios, CO is significantly enhanced, $CH_4$ is toward the high end of values found to date, $CH_3OH$ is somewhat enhanced, and $C_2H_6$ and HCN are normal to slightly enhanced. Our abundance of $CH_4$ and HCN were similar to values found in other CO-rich comets (excepting C/1996 B2, which had a lower amount of $CH_4$), while $CH_3OH$ and $C_2H_6$ have similar to enhanced abundances among these comets.

In this work, we presented abundances ratios (in percent) relative to water. If we consider the possible effect of water not being fully sublimated within our field-of-view (FOV) at the observed heliocentric distance (~ 2 AU), and adjust the water fraction to achieve "normal" abundances of $C_2H_6$ and HCN, the water production rate would increase by ~ 60–90%. Our non-detection of water in 2011 August could be related to this effect. If this added fraction were pure water ice, its addition would bring the abundance ratios of most other minor species (i.e., $C_2H_6$, $CH_4$, $CH_3OH$ and HCN) into agreement with values in the "organics-normal" group (Mumma et al. 2003; Bockelée-Morvan et al. 2004; DiSanti and Mumma 2008). However, the CO abundance would decrease to about 7.26%, which is still higher than that seen in most OC comets. Possible explanations for a high CO composition suggest accretion in the outer region of the proto-planetary disk that underwent minor (if any) thermal processing before incorporation into the nucleus. Upcoming measurements at smaller heliocentric distances will provide additional tests of this issue.

**Acknowledgments.**

We thank ESO’s VLT team. LP thanks the NASA Postdoctoral Program. MJM, GLV, and MAD acknowledge NASA's Astrobiology, PAST, and PATM. BPB acknowledges NSF. HB and ML acknowledge support from IMPRS and GIF.

Table 1. Molecular Parameters for Primary Volatiles in C/2009 P1 (Garradd)*

| Species | Time (UT) | $T_i$ [a] (min) | Lines | ν [b] ($cm^{-1}$) | $T_{rot}$ (K) | GF [c] | Global Q [d] ($10^{26}$ $s^{-1}$) | Abundance % |
|---|---|---|---|---|---|---|---|---|
| | 2011 August 7 | | † $R_h$ = 2.40 AU, Δ = 1.47 AU, P.A. = 187.9º, α = 12.6º | | | | | |
| **$H_2O$** | 5:15−6:01 | 32 | 9 | 3421.38 | (40) [e] | (1.5) [f] | < 208.9 | 100 |
| $C_2H_6$ | 6:13−7:03 | 32 | 4 | 2986.73 | (40) | (1.5) | 6.1 ± 0.7 | > 2.90 |
| | 2011 September 17 | | $R_h$ = 2.03 AU, Δ = 1.55 AU, P.A. = 98.0º, α = 28.7º | | | | | |
| **$H_2O$** | 0:25−1:15 | 40 | 7 | 3397.19 | (50) [g] | 1.8 | 841.8 ± 64.2 | 100 |
| $C_2H_6$ | 1:24−2:50 | 52 | 4 | 2988.42 | $49^{+29}_{-13}$ | 1.4 | 8.8 ± 0.7 | 1.04 ± 0.11 |
| $CH_4$ | “ | “ | 2 | 3033.63 | (50) | 1.4 | 11.3 ± 1.5 | 1.34 ± 0.21 |
| | 2011 September 18 | | $R_h$ = 2.02 AU, Δ = 1.55 AU, P.A. = 97.0º, α = 29.0º | | | | | |
| **$H_2O$** | 23:56−1:06 | 48 | 2 | 2148.01 | (50) | 1.4 | 833.1 ± 147.8 | 100 |
| CO | ” | “ | 8 | 2152.97 | 51 ± 2 | 1.4 | 113.2 ± 8.6 | 13.45 ± 1.45 [h] |
| HCN | 1:20−2:38 | 64 | 8 | 3310.38 | $52^{+9}_{-8}$ | 1.5 | 2.9 ± 0.3 | 0.35 ± 0.04 [h] |
| $C_2H_2$ | “ | “ | 9 | 3296.45 | (50) | (1.5) | < 1.1 | < 0.13 [h] |
| $NH_3$ | “ | “ | 6 | 3320.77 | (50) | (1.5) | < 14.2 | < 1.69 [h] |
| | 2011 September 19 | | $R_h$ = 2.02 AU; Δ = 1.56 AU, P.A. = 96.0º, α = 29.3º | | | | | |
| **$H_2O$** | 0:22−1:15 | 40 | 1 | 2039.96 | (50) | (1.5) | 883.3 ± 306.6 | 100 |
| OCS | “ | “ | 7 | 2057.61 | (50) | (1.5) | < 1.6 | < 0.20 [h] |
| HDO | 1:26−2:22 | 40 | 9 | 2684.78 | (50) | (1.5) | < 7.5 | < 0.89 [h,i] |
| | 2011 September 20 | | $R_h$ = 2.01 AU; Δ = 1.57 AU, P.A. = 95.1º, α = 29.5º | | | | | |
| **$H_2O$** [j] | 23:41−1:07 | 60 | 6 | 3306.7 | (50) | (1.5) | 838.7 ± 170.2 | 100 |
| HCN | “ | “ | 7 | 3305.75 | 47 ± 6 | 1.5 | 3.1 ± 0.3 | 0.37 ± 0.05 [k] |
| $C_2H_2$ | “ | “ | 8 | 3288.90 | (50) | (1.5) | < 1.2 | < 0.14 [k] |
| $NH_3$ | " | " | 6 | 3328.25 | (50) | (1.5) | < 11.8 | < 1.40 [k] |
| $CH_3OH$ | 1:15–2:33 | 56 | 9 | 2837.52 | $48^{+10}_{-6}$ | 1.7 | 32.8 ± 3.7 | 3.90 ± 0.53 [k] |
| | 2011 September 21 | | $R_h$ = 2.00 AU; Δ = 1.58 AU, P.A. = 94.2º, α = 29.8º | | | | | |
| **$H_2O$** | 23:50–0:44 | 42 | 11 | 3403.40 | $51^{+24}_{-19}$ | 1.5 | 840.5 ± 66.9 | 100 |
| CO | 0:51–1:32 | 32 | 8 | 2152.97 | 60 ± 5 | 1.7 | 97.2 ± 6.7 | 11.56 ± 1.19 |
| $C_2H_6$ | 2:02–2:34 | 24 | 4 | 2988.42 | $48^{+25}_{-15}$ | 1.4 | 8.2 ± 0.7 | 0.98 ± 0.11 |
| $CH_4$ | " | " | 2 | 3033.63 | (50) | 1.4 | 9.6 ± 2.6 | 1.14 ± 0.33 |

* Mixing ratios (in percent) are expressed relative to $H_2O$. Uncertainties represent 1σ, and upper limits represent 3σ. The reported error in production rate includes the line-by-line scatter in measured column densities, along with photon noise, systematic uncertainty in the removal of the cometary continuum, and (minor) uncertainty in rotational temperature. (For further details, please contact the corresponding author.)

† $R_h$: Heliocentric distance. Δ: Geocentric distance. P.A.: Position angle of the extended Sun-comet vector. α: Solar Phase (Sun–comet–Earth) angle; see upper-right compass in Fig. 4. (These values represent the mid-point of data acquisition.)

a. Total on-source integration time.

b. Mean frequency of all emission lines (used for this reduction) from a particular species.

c. Growth factor.

d. Global production rate, after applying a measured growth factor (and an error of ± 0.1) to the nucleus-centered (NC) production rate.

e. Assumed rotational temperature based on September observations.

f. For molecules whose growth factor is not measured directly, we adopt GF = (1.5).

g. We tabulate the retrieved $T_{rot}$ and confidence limits. Measured temperatures are consistent with 53 K for all molecules. We adopted $T_{rot}$ = 50 K when calculating the NC production rates for all species whose $T_{rot}$ determination was not possible. This is indicated as (50).

h. The stated molecular abundance ratios use the water production rate from September 17.

i. D/H in water < 29 VSMOW (3σ).

j. Retrieval of the water production rate is based on OH* emission lines, a direct proxy for water (Bonev et al. 2006).

k. The stated molecular abundance ratios use the water production rate from September 21.

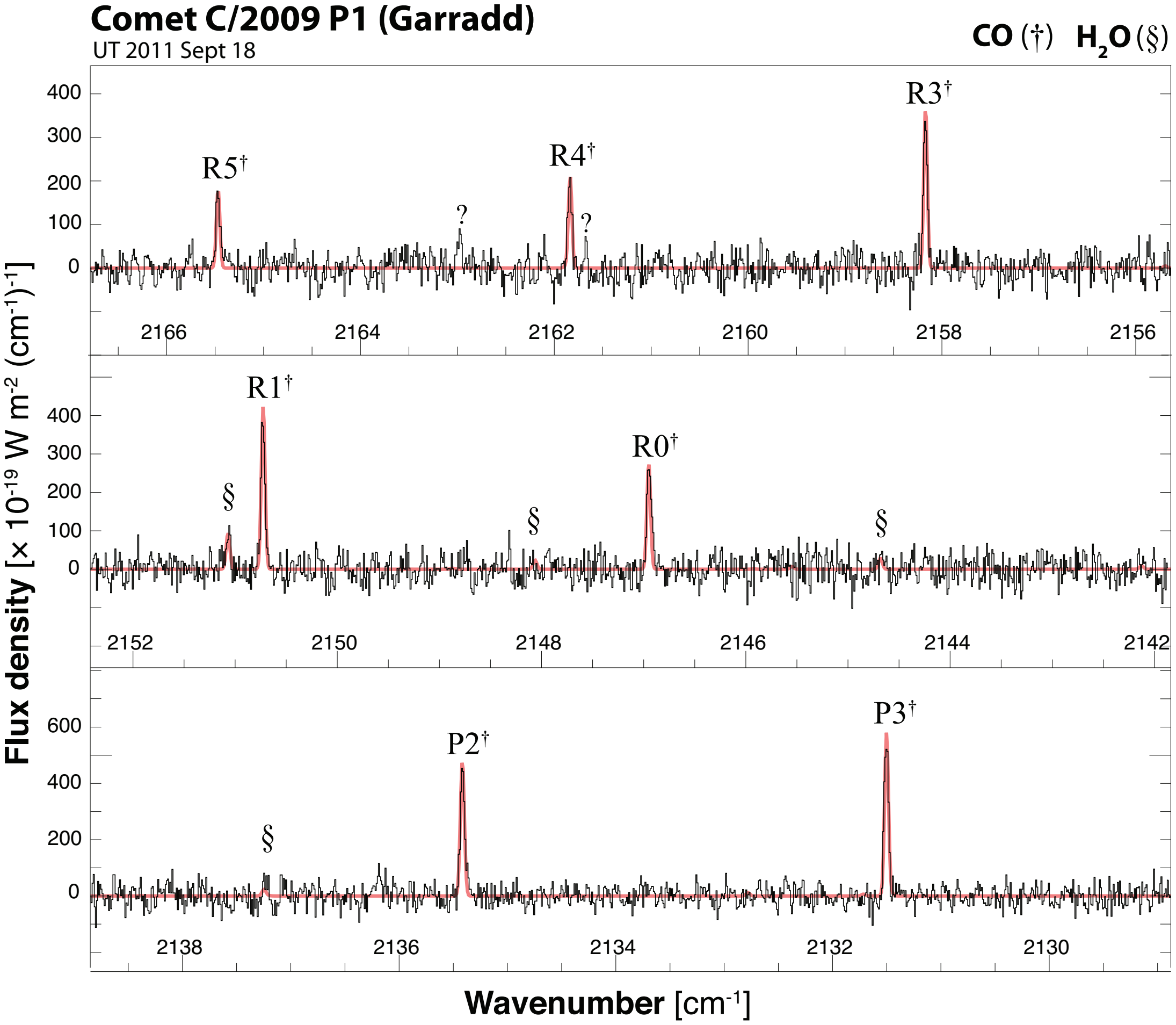

Figure 1. Emission spectra of C/2009 P1 near 4.7 µm. Ro-vibrational lines of the CO fundamental band (v = 1–0) and $H_2O$ (hot-bands) are apparent. The CO analysis was based on the CO R6–P3 lines shown here, along with the R7 line (not shown). Here, and in Fig. 2, the continuous (underlying) red line depicts the modeled spectra. Also, several unidentified lines are indicated ('?').

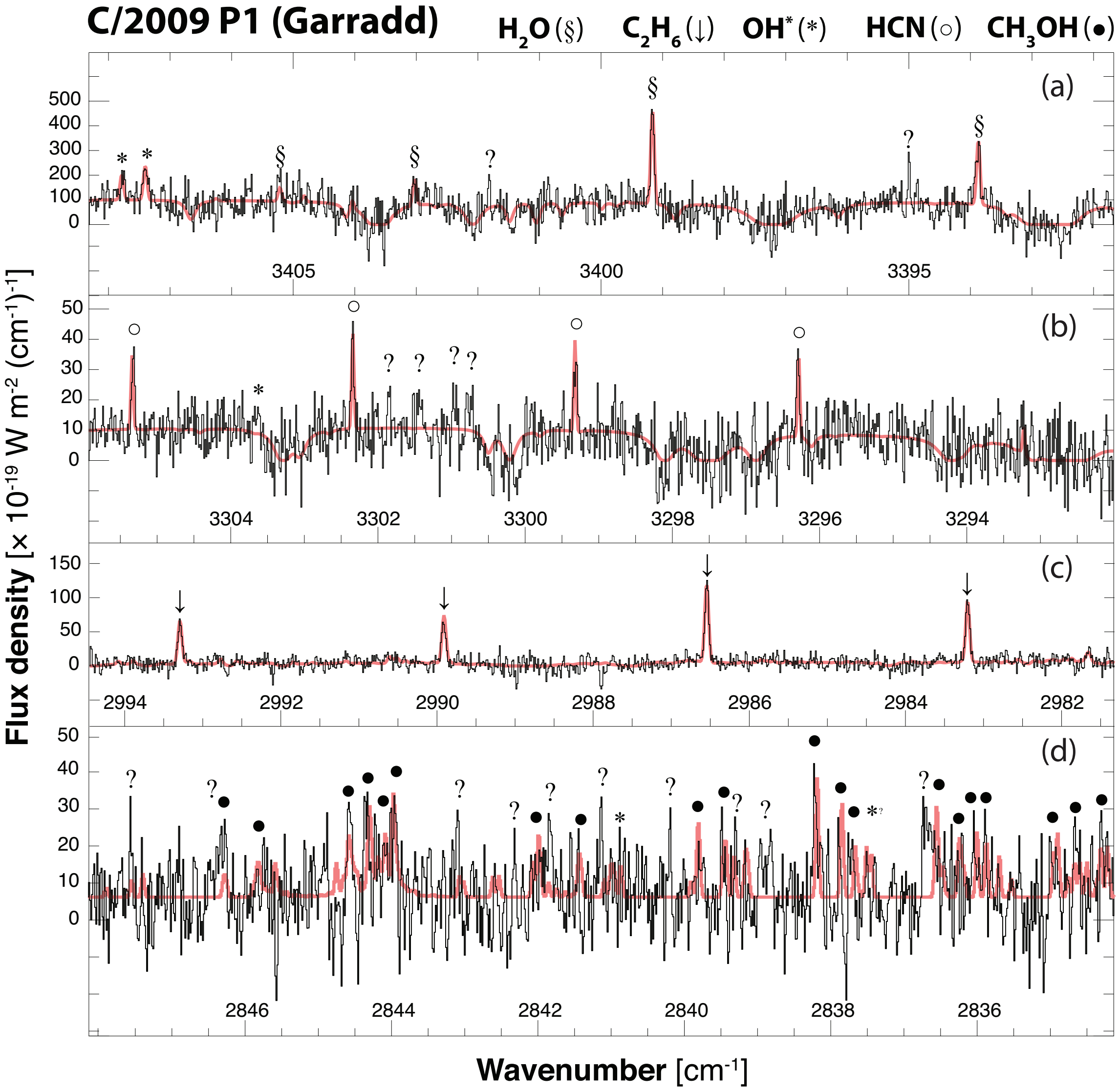

Figure 2. Detections of four primary volatiles and OH prompt emission (a direct proxy for water) in comet C/2009 P1 (Garradd). The continuum from cometary dust is detected. (a) $H_2O$ and OH* lines detected on UT 2011 Sept 17. (b) HCN lines detected on UT 2011 Sept 18. (c) Ethane $\nu_7$ Q-branches detected on UT 2011 Sept 17. (d) Methanol emission lines ($\nu_3$-band) detected on UT 2011 Sept. 20.

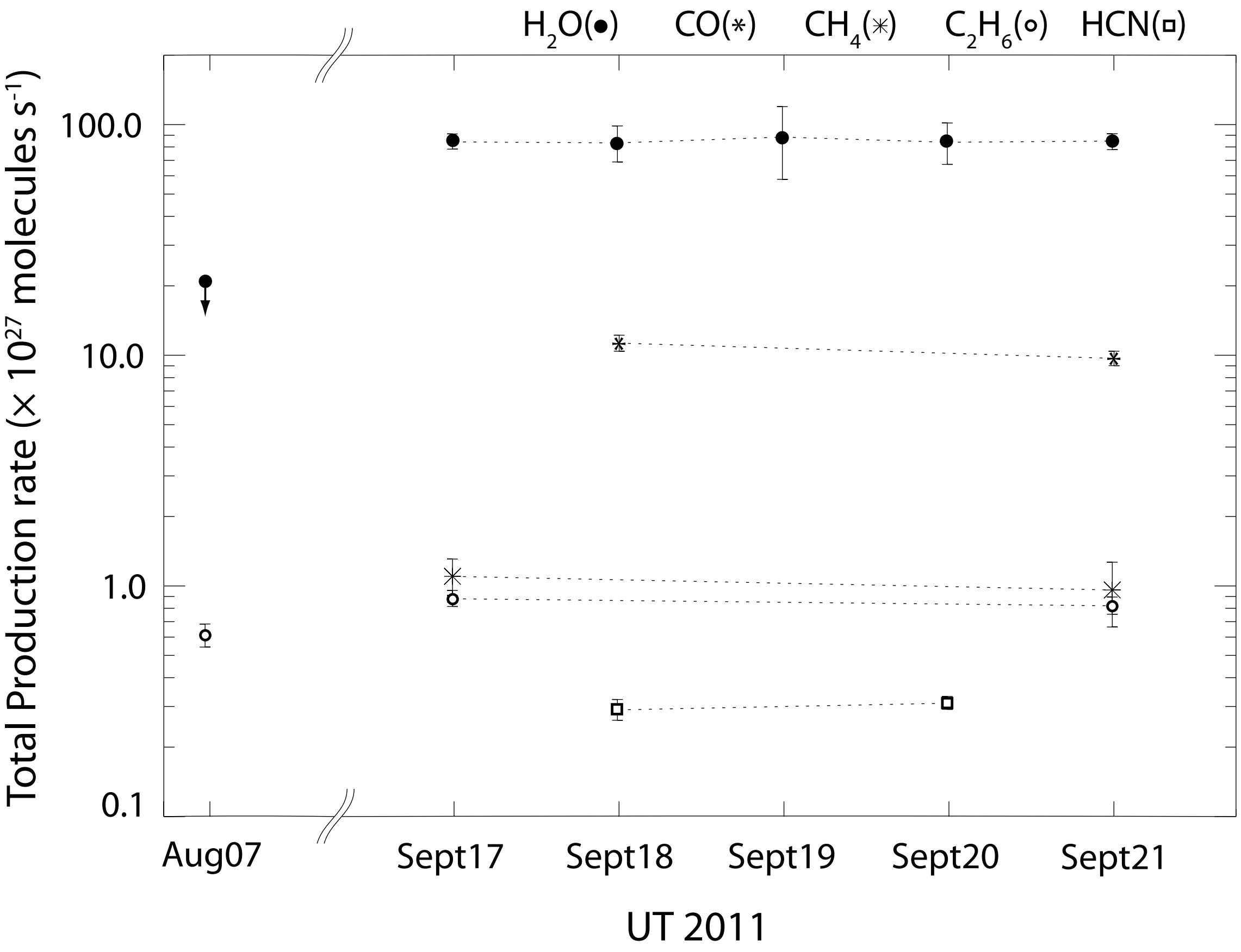

Figure 3. Evolution of total production rates during our observing campaign.

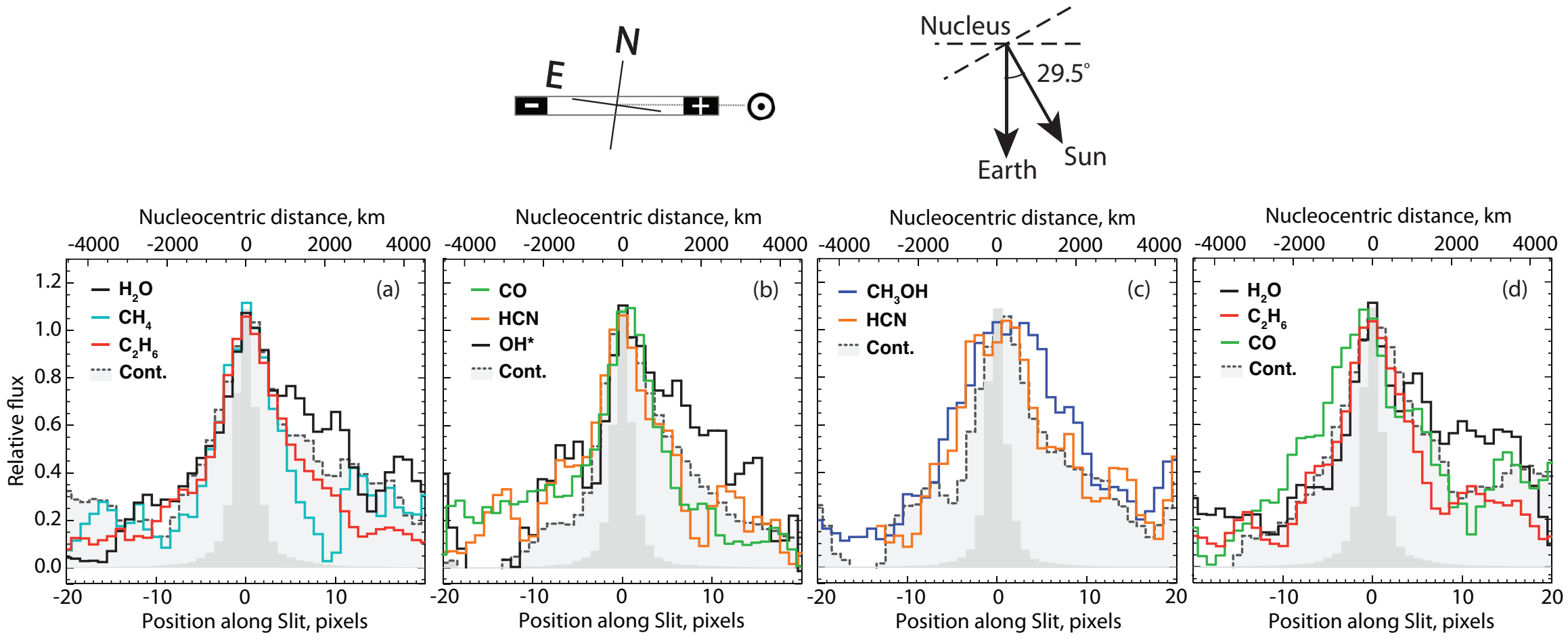

Figure 4. Spatial profiles of primary volatiles and continuum in comet C/2009 P1 on four nights: (a) UT 2011 Sept 17, (b) Sept 18, (c) Sept 20, (d) Sept 21. The slit was positioned along the projected comet-Sun radius vector (P.A. ~ 96º), and the projected sunward direction (+) and solar phase angle (29.5º) are marked. The measured stellar Point-Spread-Function and continuum are shaded (dark grey and light grey, respectively). Profiles are discussed in Section 3.3.